\documentclass{article}
\usepackage{spconf,amsmath,graphicx,amssymb,bm,pgfplots,comment,tikz,enumitem,hyperref}
\usepackage[caption=false]{subfig}
\pgfplotsset{compat=1.14}
\pgfmathsetseed{1}
\usetikzlibrary{patterns}

\DeclareMathOperator{\Tr}{Tr}

\title{AUDIO-VISUAL CALIBRATION WITH POLYNOMIAL REGRESSION FOR 2-D PROJECTION USING SVD-PHAT}
\name{Fran\c{c}ois Grondin\thanks{This work was supported by the Toyota Research Institute.}, Hao Tang, James Glass}
\address{Computer Science and Artificial Intelligence Laboratory\\
Massachusetts Institute of Technology\\
Cambridge, MA 02139, USA \\
\small\texttt{\{fgrondin, haotang, glass\}@mit.edu}}

\begin{document}
%
\maketitle
\begin{abstract}
This paper proposes a straightforward 2-D method to spatially calibrate the visual field of a camera with the auditory field of an array microphone by generating and overlaying an acoustic image over an optical image. Using a low-cost microphone array and an off-the-shelf camera, we show that polynomial regression can deal efficiently with non-linear camera distortion, and that a recently proposed sound source localization method for real-time processing, SVD-PHAT, can be adapted for this task.
\end{abstract}
\section{Introduction}
\label{sec:intro}

The cocktail party effect makes distant speech processing a challenging task for automatic speech recognition (ASR) systems \cite{tang2018study}.
A beamformer with multiple microphones is therefore often used to enhance the corrupted speech signal \cite{heymann2015blstm,lee2019deep,sun2018effect}.
Some beamforming methods (e.g. delay and sum and the minimum variance distortionless response (MVDR) \cite{habets2010new}) rely on the target source direction of arrival (DOA).

Sound source localization is based on estimating the DOA, and can rely on high resolution methods such as Multiple Signal Classification (MUSIC) \cite{schmidt1986multiple}, or Steered-Response Power Phase Transform (SRP-PHAT) \cite{dibiase2001robust} methods.
MUSIC was initially used for narrowband signals, and subsequently adapted to broadband signals to make localization robust to additive noise \cite{ishi2009evaluation}.
However, this method requires performing online eigenvalue value decomposition, which involves a significant amount of computation, and makes real-time processing challenging.
Alternatively, SRP-PHAT relies on the Generalized Cross-Correlation with Phase Transform (GCC-PHAT) between each pair of microphones, and can be computed with low-cost embedded hardware \cite{grondin2019lightweight}.
We recently proposed the Singular Value Decomposition with Phase Transform (SVD-PHAT) \cite{grondin2019svd, grondin2019multiple} that reduces the amount of computations typically involved in the exact SRP-PHAT solution, while preserving accuracy.

It is often desirable to not only localize the exact source position in space, but project a heat map of the estimated sound source energy on the corresponding video image, for multimodal classification and segmentation \cite{noda2015audio, ephrat2018looking}.
There are acoustic cameras on the market, but these usually have a large aperture, use numerous microphones, and are calibrated using a cumbersome procedure that involves specialized hardware, which makes them less suitable for robotics applications \cite{odonovan2007microphone}.
Cech proposes an approach that maps sound sources on an image using linear regression for the Nao robot, but their method requires a stereoscopic camera \cite{cech2013active}.
Deleforge investigates DOA estimation and projection on an image, but their approach is limited to a binaural setup \cite{deleforge2015colocalization}.

This paper proposes a simple yet efficient method to calibrate a low-cost camera and a planar microphone array with an arbitrary shape to generate an acoustic image.
The proposed approach uses a polynomial regression to deal with camera distortion and demonstrates that it is possible to 1) calibrate the acoustic camera without any specific hardware (e.g. loudspeakers, checkerboard, etc.) and using few measurements 2) adapt the SVD-PHAT method to generate efficiently an acoustic image in real-time with few computations.

\section{Task definition}

We define our goal as the ability to generate an acoustic image to overlay on a corresponding optical image.
In the current setup, a microphone array is equipped with $M$ microphones, indexed by $m \in \mathcal{M} = \{1, \dots, M\}$, with a camera positioned in the center of the array.
We assume the camera provides an optical image with a width of $U$ pixels and a height of $V$ pixels.
The task then is to calculate the sound phase transform energy for each pixel as in (\ref{eq:task_acimg}), such that the acoustic image can be overlaid on the optical image, where $\mathcal{U} = \{1,2,\dots,U\}$ and $\mathcal{V} = \{1,2,\dots,V\}$.
\begin{equation}
    f: \mathcal{U}\times\mathcal{V} \rightarrow \mathbb{R}
    \label{eq:task_acimg}
\end{equation}

To produce an accurate acoustic image properly aligned with the optical image, the function $f$ therefore needs to handle the camera distortion and the phase shift between each pair of microphones.
This phase shift usually depends on the propagation of sound in air given the direction of arrival of the sound source, but can also depend on the sound card hardware calibration.

\section{Proposed method}
\label{sec:proposed}

To solve the alignment task, one approach would be to map the vector of $M$ pairs of TDOAs to a coordinate in the unit sphere, followed by mapping the coordinate on the unit sphere to the plane coordinate projected according to the camera parameters.
However, there are several drawbacks with this approach:
1) the camera parameters are not always available,
2) two calibration procedures, one for the camera and one for the microphone array, are needed, and finally, 
3) higher calibration accuracy is required as the error propagates through the two projections.
In this paper, we take a more direct approach, mapping each vector of $M$ pairs of TDOAs to a coordinate on the display, thereby avoiding projecting the coordinates on the unit sphere and the calibration of the camera.

Formally, we consider only the steered response energy with the phase transform, such that each pixel $(u,v)$ is assigned a time difference of arrival (TDOA) $\tau_{u,v,i,j}$ for each microphone pair $(i,j) \in \mathcal{P} = \{(x,y) \in \mathcal{M}^2: x < y\}$:
\begin{equation}
    g: \mathcal{U} \times \mathcal{V} \rightarrow \mathcal{T}^{P}
    \label{eq:task_tdoas}
\end{equation}
where $\mathcal{T} \in [-\tau_{max},+\tau_{max}]$, and $P = |\mathcal{P}| = M(M-1)/2$ corresponds to the number of pairs of microphones.
The maximum TDOA (in samples) $\tau_{max}$ between microphones $i$ and $j$ corresponds to:
\begin{equation}
    \tau_{max} = (1+\rho)\max_{(i,j) \in \mathcal{P}}{\left\{\left(f_S/c\right)\lVert \mathbf{r}_i - \mathbf{r}_j \rVert_2\right\}}
\end{equation}
where $\rho \in \mathbb{R}_+$ is a user-defined parameter (here chosen to be $0.1$) that allows overestimating the TDOA to deal with potential disparities between the free-field sound propagation model and the experimental setup.
The vectors $\mathbf{r}_i$ and $\mathbf{r}_j$ stand for the $xyz$-positions of the microphones $i$ and $j$ (in meter), and $\lVert\dots\rVert_2$ stands for the $l^2$-norm.
The constants $f_S \in \mathbb{R}_+$ and $c \in \mathbb{R}_+$ hold the sample rate (in samples/second) and the speed of sound (in meter/second), respectively.

We first find a function $g$, which predicts the TDOAs of all pairs of microphones for each pixel $(u,v)$ in the image.
We propose using a polynomial regression since its nonlinearity and small number of parameters make it suitable to model the optical and acoustic phenomena with few measurements:
\begin{equation}
    \tau_{u,v,i,j} = \sum_{a=0}^{L}\sum_{b=0}^{L}x(u)^{a}y(v)^{b}c_{a,b,i,j}
    \label{eq:proposed_polyreg_tau}
\end{equation}
where $c_{a,b,i,j}$ is one of the $(L+1)^2$ parameters for the microphone pair $(i,j)$.
For numerical stability, the position in pixels is normalized as follows:
\begin{equation}
    \begin{array}{c}
    x(u) = (2u-U-1)/(U-1) \\
    y(v) = (2v-V-1)/(V-1)
    \end{array}
\end{equation}

\begin{figure}[!ht]
    \centering
    \includegraphics[width=0.7\columnwidth]{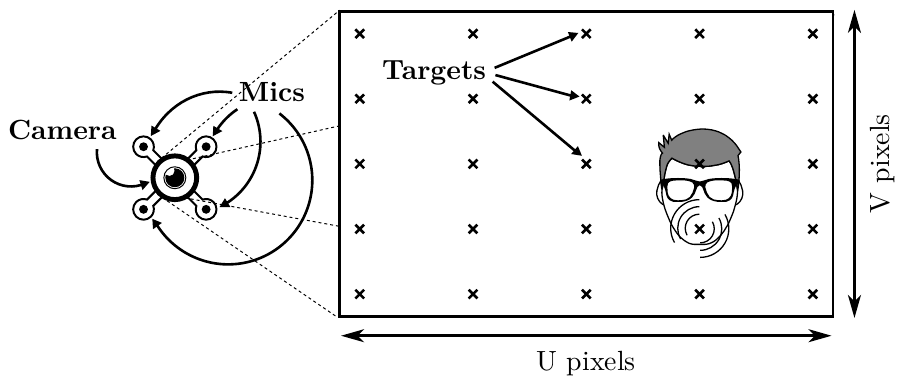}
    \caption{Calibration with a sound source. We sequentially position a sound source that generates white noise. For this application, the sound source may consist of a human speaker that produces a fricative sound such as the phoneme \texttt{/s/}, which is more convenient than using a loudspeaker that plays white noise.}
    \label{fig:proposed_polyreg_calibration}
\end{figure}

To estimate each coefficient $c_{a,b,i,j}$, we sample $T$ pixels on the image, that we call targets, as shown in Figure \ref{fig:proposed_polyreg_calibration}.

We denote each target by $(u_t,v_t) \in \mathcal{U} \times \mathcal{V}$, for $t \in \{1, \dots, T\}$.
The regression introduced in (\ref{eq:proposed_polyreg_tau}) can be expressed in matrix form as follows:
\begin{equation}
    \bm{\tau} = \mathbf{A}\mathbf{c}
    \label{eq:proposed_polyreg_matrix}
\end{equation}
where the matrix $\mathbf{A} \in \mathbb{R}^{T \times (L+1)^2}$ holds the polynomials obtained from the target positions, as shown in (\ref{eq:proposed_polyreg_Am}).
\begin{equation}
    \mathbf{A} = \left[
    \begin{array}{ccc}
    x(u_1)^0 y(v_1)^0 & \dots & x(u_1)^L y(v_1)^L \\
    \vdots & \ddots & \vdots \\
    x(u_T)^0 y(v_T)^0 & \dots & x(u_T)^L y(v_T)^L \\
    \end{array}
    \right]
    \label{eq:proposed_polyreg_Am}
\end{equation}

For each target $t$, we compute the TDOAs for all pairs of microphones ($\tau_{u_t,v_t,1,2}, \tau_{u_t,v_t,1,3}, \dots, \tau_{u_t,v_t,M-1,M}$) using GCC-PHAT.
All TDOAs are then concatenated to form the matrix $\bm{\tau} \in \mathcal{T}^{T \times P}$, as shown in (\ref{eq:proposed_polyreg_taum}).
\begin{equation}
    \bm{\tau} = \left[
    \begin{array}{ccc}
    \tau_{u_1,v_1,1,2} & \dots & \tau_{u_1,v_1,M-1,M} \\
    \vdots & \ddots & \vdots \\
    \tau_{u_T,v_T,1,2} & \dots & \tau_{u_T,v_T,M-1,M} \\
    \end{array}
    \right]
    \label{eq:proposed_polyreg_taum}
\end{equation}

The matrix $\mathbf{c} \in \mathbb{R}^{(L+1)^2 \times P}$ then holds all the regression parameters, and it is possible to recover these parameters with the expression in (\ref{eq:proposed_polyreg_inv}).
\begin{equation}
    \mathbf{c} = \mathbf{A}^{+} \bm{\tau}
    \label{eq:proposed_polyreg_inv}
\end{equation}
with the Moore-Penrose inverse $\mathbf{A}^{+} = (\mathbf{A}^T\mathbf{A})^{-1}\mathbf{A}^T \in \mathbb{R}^{(L+1)^2 \times T}$, where $\{\dots\}^T$ stands for the matrix transpose.
Note that the number of targets needs to be greater than the lower bound $T \geq (L+1)^2$ to make the system overdetermined.

The TDOAs $\tau_{u,v,i,j}\ \forall (i,j)\ \in \mathcal{P}$ can now be predicted for each pixel $(u,v)$ using the polynomial regression model in (\ref{eq:proposed_polyreg_tau}).
The SVD-PHAT method then generates the projection matrix $\mathbf{W}\in\mathbb{C}^{UV\times P(N/2+1)}$:
\begin{equation}
\mathbf{W} = \left[
\begin{array}{ccc}
    W_{1,1,1,2}(0) & \cdots & W_{1,1,M-1,M}(N/2) \\
    \vdots & \ddots & \vdots \\
    W_{U,V,1,2}(0) & \cdots & W_{U,V,M-1,M}(N/2) \\
\end{array}
\right]
\end{equation}
where each element $W_{u,v,i,j}(f)\in\mathbb{C}$ corresponds to:
\begin{equation}
    W_{u,v,i,j}(f) = \exp(2\pi\sqrt{-1}f\tau_{u,v,i,j}/N)
\end{equation}
where $N$ stands for the number of samples per frame, and $f \in \{0, 1, \dots, N/2\}$ corresponds to the Short-Time Fourier Transform (STFT) frequency bin index.
The frames from all microphones are concatenated in a supervector $\mathbf{X}\in\mathbb{C}^{P(N/2+1)\times 1}$:
\begin{equation}
    \mathbf{X} = [
    \begin{array}{cccc}
    X_{1,2}(0) & X_{1,2}(1) & \dots & X_{M-1,M}(N/2)
    \end{array}
    ]
\end{equation}
where each element $X_{i,j}(f) \in \mathbb{C}$ corresponds to the normalized cross-correlation according to the phase transform:
\begin{equation}
    X_{i,j}(f) = X_i(f)X_j(f)^*/|X_i(f)X_j(f)|
\end{equation}
and $\{\dots\}^*$ stands for the complex conjugate.
At this point, the phase transform energy for each point $(u,v)$ can be displayed on a $U\times V$ acoustic image, reconstructed from the vector $\mathbf{Y} \in \mathbb{R}^{UV \times 1}$:
\begin{equation}
    \mathbf{Y} = \left[
        \begin{array}{ccc}
            Y_{(1,1)} & \dots & Y_{(U,V)}
        \end{array}
    \right]^T = \Re\{\mathbf{W}\mathbf{X}\}
    \label{eq:proposed_svdphat_brute}
\end{equation}

The initial matrix multiplication $\mathbf{W}\mathbf{X}$ is of complexity $\mathcal{O}(UVPN)$, which involves a significant amount of computation, and makes real-time difficult to achieve.
To reduce this complexity, we decompose the supermatrix $\mathbf{W}$ as follows using singular value decomposition (SVD):
\begin{equation}
    \mathbf{W} \approx \mathbf{U}\mathbf{S}\mathbf{V}^H
\end{equation}
where the rank $K$ is chosen to achieve accurate reconstruction, according to the following condition:
\begin{equation}
    \Tr{\{\mathbf{S}\mathbf{S}^T\}} \geq (1 - \delta)\Tr{\{\mathbf{W}\mathbf{W}^H\}}
    \label{eq:proposed_svdphat_condition}
\end{equation}
with $\Tr\{\cdot\}$ that denotes the trace operator, and $\delta\in[0,1]$ being a user-defined parameter, chosen to be $10^{-5}$ as in \cite{grondin2019svd}.
The acoustic image can thus be computed as follows:
\begin{equation}
    \mathbf{Y} = \Re\{(\mathbf{U}\mathbf{S})(\mathbf{V}^H\mathbf{X})\}
    \label{eq:proposed_svdphat_optimized}
\end{equation}
with $\Re\{\dots\}$ that extracts the real part.
This then reduces the complexity to $\mathcal{O}(UVK + KPN)$.

\section{Results}
\label{sec:results}

\begin{figure}[!ht]
    \centering
    \includegraphics[width=0.6\columnwidth]{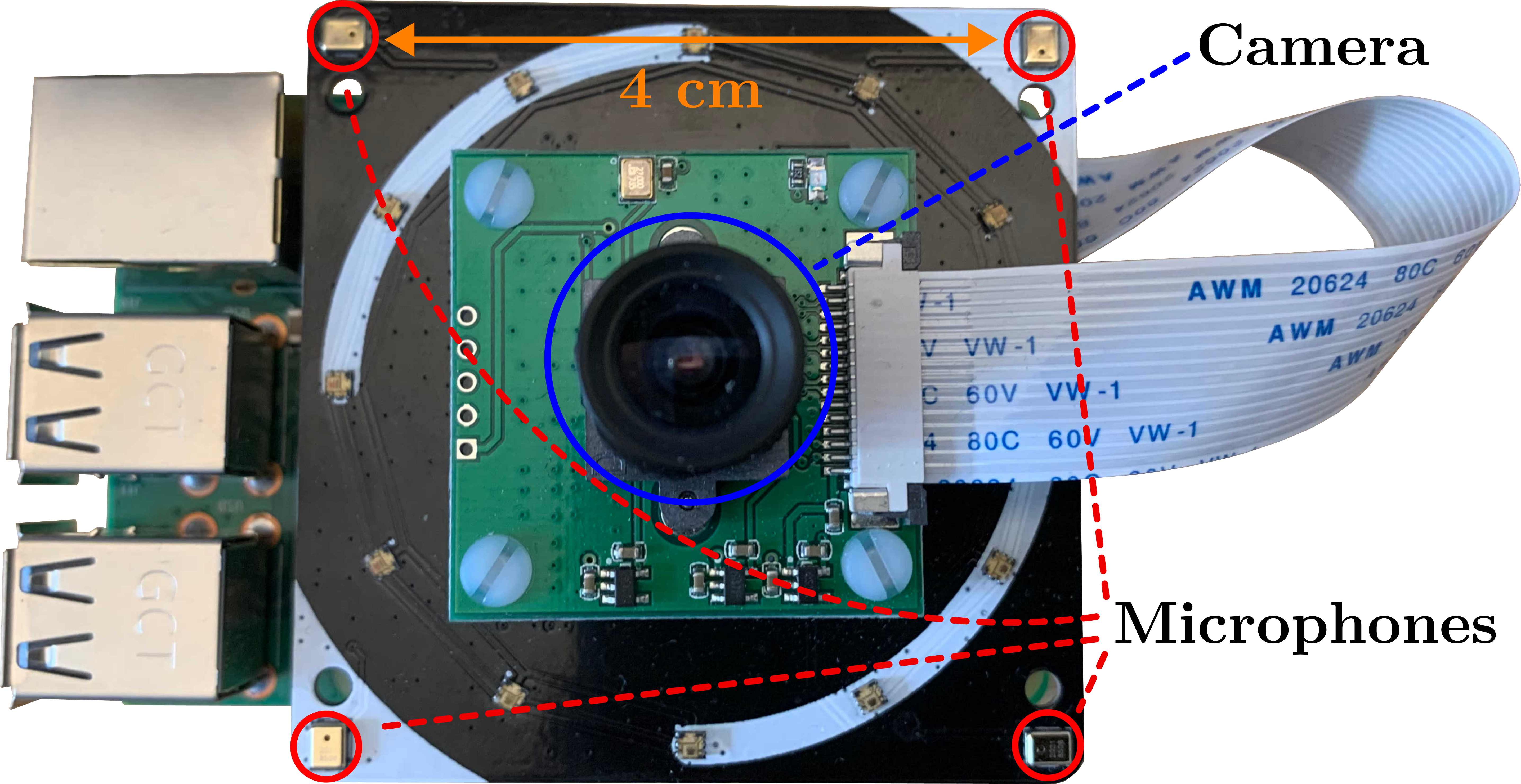}
    \caption{ReSpeaker 4-mic array with a Arducam 5 Megapixels 1080p Sensor OV5647 Mini Camera mounted in the middle}
    \label{fig:results_setup}
\end{figure}

Simulations are first performed to validate the accuracy of the polynomial regression to estimate TDOAs associated with each pixel once the image is distorted by the camera.
We then show acoustic images generated with the proposed polynomial regression method and SVD-PHAT using low-cost off-the-shelf microphone array (e.g. a ReSpeaker 4-microphone planar array \footnote{\scriptsize\url{https://respeaker.io}}) and camera (e.g. a Arducam 5 Megapixels 1080p Sensor OV5647 Mini Camera \footnote{\scriptsize\url{https://raspberrypi.org/documentation/hardware/camera/}}), as shown in Figure \ref{fig:results_setup}.
The image resolution corresponds to the QVGA format \cite{mudigoudar2009video} ($U=320$ and $V=240$) to lower the bandwidth when transmitting the video stream over the network.
The audio sample rate is set to $f_S = 16000$ samples/sec, and we assume the speed of sound at room temperature to be $c = 343$ m/sec.

\subsection{Simulations}

Camera distortion consists of mapping $xyz$-coordinates in space to the pixels on the image as follows:
\begin{equation}
    (\hat{u},\hat{v}) = h_c(\hat{x},\hat{y},\hat{z})
    \label{eq:results_simulations_camera}
\end{equation}

We use a simple distortion model (such as the one in \cite{drap2016exact}) with one radial distortion parameter denoted by $k$ in (\ref{eq:results_simulations_uv}).
\begin{equation}
    \begin{array}{c}
        \hat{u} = (U/2)\left(\hat{x}/\hat{z}\right)\left(1+k\left(\hat{x}^2+\hat{y}^2\right)/\hat{z}^2\right) + (U/2) \\
        \hat{v} = (V/2)\left(\hat{y}/\hat{z}\right)\left(1+k\left(\hat{x}^2+\hat{y}^2\right)/\hat{z}^2\right) + (V/2)
    \end{array}
    \label{eq:results_simulations_uv}
\end{equation}

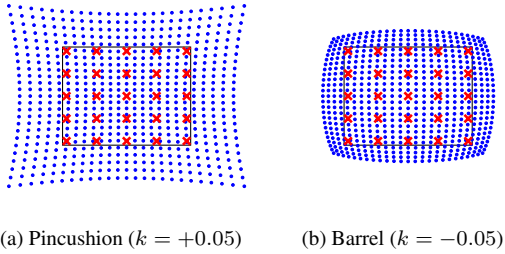
\begin{figure}[!ht]
    \centering
    \subfloat[Pincushion ($k=+0.05$)]{%
    \begin{tikzpicture}
    \begin{axis}[width=0.58\columnwidth, 
                 xticklabels={,,}, yticklabels={,,},
                 xtick style={draw=none}, ytick style={draw=none}, 
                 axis line style={transparent},
                 xmin=-160, xmax=+480, ymin=-130, ymax=+370]
        \addplot[only marks,blue,mark=*, mark size=0.5pt] table [x=U, y=V, col sep=comma] {distortion_pincushion.csv};
        \addplot[only marks,red,mark=x, mark size=2pt, line width=1pt] table [x=U, y=V, col sep=comma] {targets.csv};
        \draw (0,0) -- (320,0) -- (320,240) -- (0,240) -- cycle;
    \end{axis}
    \end{tikzpicture}
    } 
    \subfloat[Barrel ($k=-0.05$)]{%
    \begin{tikzpicture}
    \begin{axis}[width=0.58\columnwidth,
                 xticklabels={,,}, yticklabels={,,},
                 xtick style={draw=none}, ytick style={draw=none}, 
                 axis line style={transparent},
                 xmin=-160, xmax=+480, ymin=-130, ymax=+370]
        \addplot[only marks,blue,mark=*, mark size=0.5pt] table [x=U, y=V, col sep=comma] {distortion_barrel.csv};
        \addplot[only marks,red,mark=x, mark size=2pt, line width=1pt] table [x=U, y=V, col sep=comma] {targets.csv};
        \draw (0,0) -- (320,0) -- (320,240) -- (0,240) -- cycle;
    \end{axis}
    \end{tikzpicture}
    }
    \caption{Camera distortion with points projected on the image (blue circles), and the targets for calibration (red crosses).}
    \label{fig:results_simulation_distortion}
\end{figure}

Figure \ref{fig:results_simulation_distortion} shows the pincushion and barrel distortions, which depend on the parameter $k$.
The rectangle shows the receptive field of the camera, which corresponds to the image in pixels, and the red crosses stand for the $T$ targets to be sampled during calibration.
We use the free-field assumption to simulate the TDOA for each pair of microphone given the $xyz$-position of the sound source:
\begin{equation}
    (\tau_{1,2}, \tau_{1,3}, \dots, \tau_{M-1,M}) = h_s(\hat{x},\hat{y},\hat{z})
    \label{eq:results_simulations_acoustic}
\end{equation}

We then proceed as follows:
\begin{enumerate}[itemsep=-2pt]
    \item Generate a $5 \times 5$ grid of targets in the $U \times V$ image.
    \item Find the corresponding $xyz$ points with (\ref{eq:results_simulations_camera}).
    \item Compute the targets TDOAs with (\ref{eq:results_simulations_acoustic}).
    \item Perform polynomial regression with (\ref{eq:proposed_polyreg_inv}).
    \item Generate $Q$ points on a plane facing the camera denoted as $(\hat{x}_q,\hat{y}_q,\hat{z}_q)$.
    \item Project these points on the $U \times V$ image (\ref{eq:results_simulations_camera}).
    \item Estimate the TDOAs with (\ref{eq:proposed_polyreg_matrix}), as $g(h_c(\hat{x}_q,\hat{y}_q,\hat{z}_q))$.
    \item Compute the TDOAs with (\ref{eq:results_simulations_acoustic}), named $h_{s}(\hat{x}_q,\hat{y}_q,\hat{z}_q)$.
\end{enumerate}

Finally, we compute the Root Mean Square Error (RMSE), between the estimated TDOAs and exact ones obtained with the acoustic model, as shown in (\ref{eq:results_simulations_rmse}).
We initially generate $Q = 1,000,000$ points, but the final number varies with $k$ as we only keep points that are projected inside the $U \times V$ image.
\begin{equation}
    RMSE = \frac{1}{QP}\sum_{q=1}^{Q}{\lVert h_{s}(\hat{x}_q,\hat{y}_q,\hat{z}_q) - g_{\tau}(h_{c}(\hat{x}_q,\hat{y}_q,\hat{z}_q)) \rVert_2}
    \label{eq:results_simulations_rmse}
\end{equation}

\begin{figure}[!ht]
    \centering
    \begin{tikzpicture}
    \begin{axis}[xlabel=$k$, 
                 xmin=-0.05,
                 xmax=0.05,
                 xticklabel style={/pgf/number format/fixed, /pgf/number format/precision=2},
                 scaled ticks=false,
                 ylabel=RMSE, 
                 yticklabel style={/pgf/number format/fixed, /pgf/number format/precision=2, /pgf/number format/fixed zerofill}, 
                 legend columns=2,
                 height=0.45\columnwidth, 
                 width=\columnwidth]
        \addplot[blue,mark=*] table [x=k, y=L1, col sep=comma] {rmse.csv};
        \addplot[red,mark=square*] table [x=k, y=L2, col sep=comma] {rmse.csv};
        \addplot[green,mark=triangle*] table [x=k, y=L3, col sep=comma] {rmse.csv};
        \addplot[black,mark=diamond*] table [x=k, y=L4, col sep=comma] {rmse.csv};
        \legend{$L=1$\\$L=2$\\$L=3$\\$L=4$\\}
    \end{axis}
    \end{tikzpicture}
    \caption{RMSE between the estimated and theoretical TDOAs}
    \label{fig:results_simulations_rmse}
\end{figure}
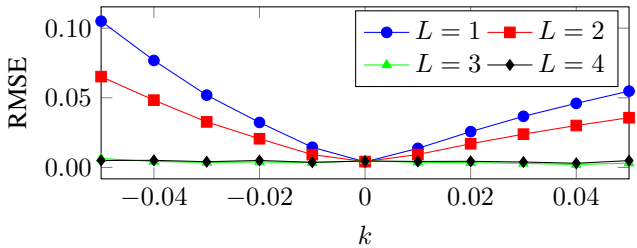

Figure \ref{fig:results_simulations_rmse} shows the results for different polynomial order $L$ while varying the distortion parameter $k$.
It is interesting to note that both linear ($L=1$) and quadratic ($L=2$) regressions fail to accurately model the TDOAs, whereas higher order regressions perform well under both barrel and pincushion distortions.
As expected, all regressions work well when there is no distortion ($k=0$).

\subsection{Experiments}

\begin{figure}[!ht]
    \centering
    \subfloat[Calibration]{\includegraphics[width=0.42\columnwidth]{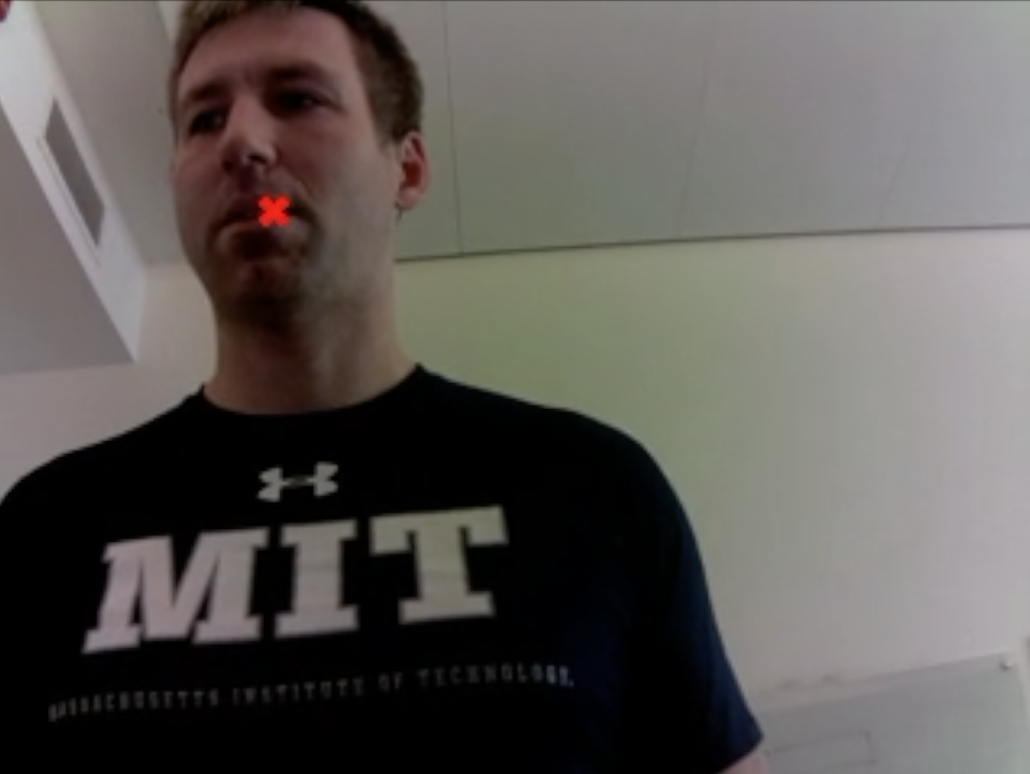} \label{fig:results_experiments_acimgs_calibration}}
    \hspace{5pt}
    \subfloat[Speech]{\includegraphics[width=0.42\columnwidth]{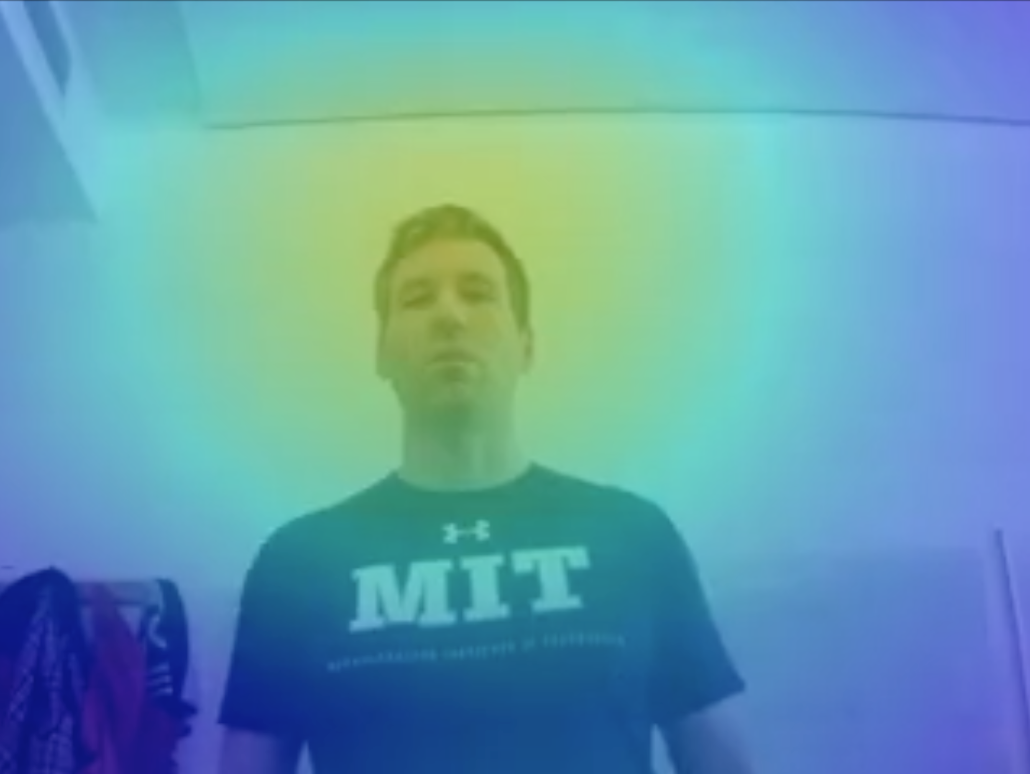} \label{fig:results_experiments_acimgs_speech}} \\
    \subfloat[Smartphone playing music]{\includegraphics[width=0.42\columnwidth]{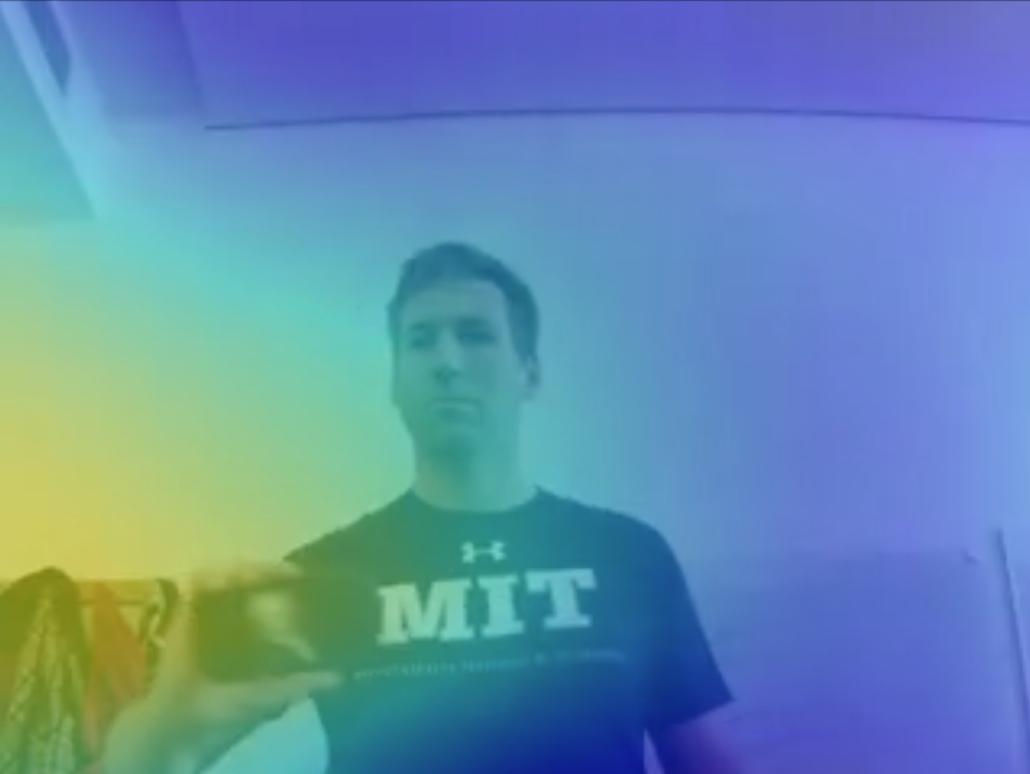} \label{fig:results_experiments_acimgs_phone}}
    \hspace{5pt}
    \subfloat[Finger snapping]{\includegraphics[width=0.42\columnwidth]{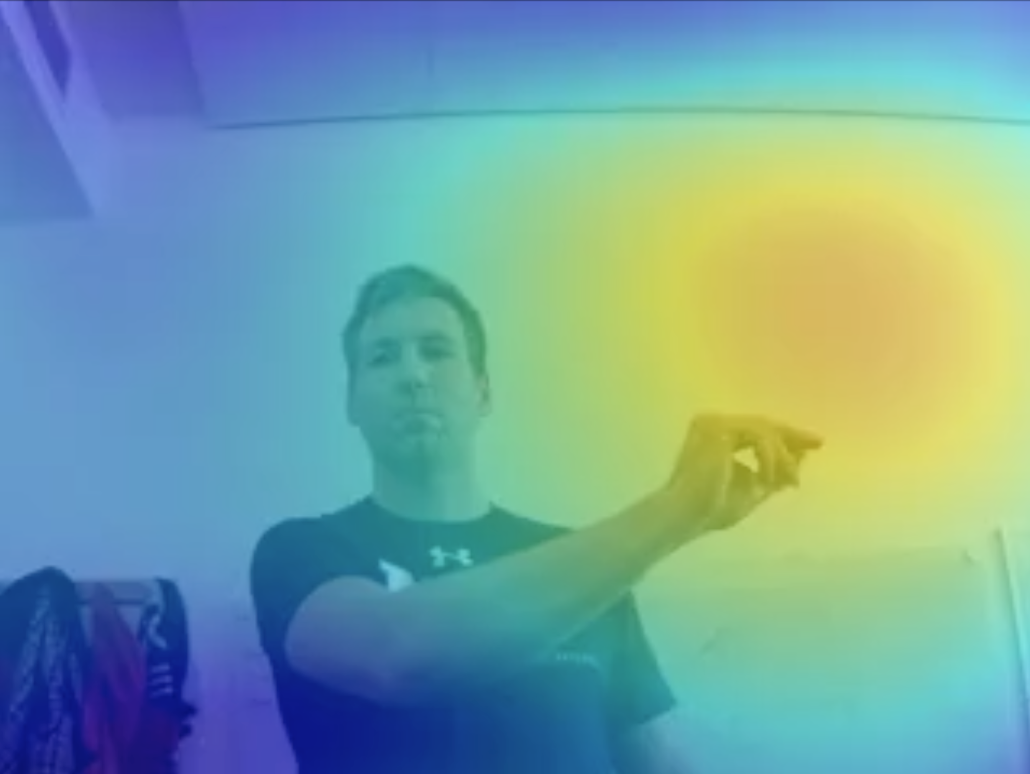} \label{fig:results_experiments_acimgs_finger}}    
    \caption{Calibration of the microphone array with camera, and generated acoustic image obtained with SVD-PHAT.}
    \label{fig:results_experiments_acimgs}
\end{figure}

Experiments are performed with the setup shown in Figure \ref{fig:results_setup}.
Calibration is performed with $T=25$ targets that sample the image as shown in Figure \ref{fig:results_experiments_acimgs_calibration} and a polynomial regression of order $L=3$.
A target is overlaid on the optical image, and the user positions their lips using visual feedback from the computer monitor and produces sound.
Once the sound is detected, the TDOAs are computed and the next target is displayed on the screen.
The SVD-PHAT method is then used to overlay the acoustic image on the optical image captured by the camera.
With this setup, and a frame size of $N=512$ samples, the total number of complex multiplications would equal $UVP(N/2+1) = 118,425,600$ when using (\ref{eq:proposed_svdphat_brute}), whereas this number reduces to $UVK + KP(N/2+1) = 2,506,944$ with SVD-PHAT when $K=32$ (to satisfy the condition in (\ref{eq:proposed_svdphat_condition})) when using (\ref{eq:proposed_svdphat_optimized}).
The reduction by a factor of $47$ makes it possible to generate the acoustic image in real-time on a laptop with an Intel Core i5 processor clocked at 2.8 GHz.
Figures \ref{fig:results_experiments_acimgs_speech}-\ref{fig:results_experiments_acimgs_finger} show the acoustic image generated by speech, a smartphone playing music and finger snapping.
Results demonstrate the effectiveness of the method.
The center of mass of the heat map is sometimes slightly shifted, but this is mostly due to the small aperture of the array which makes localization sensitive to a small phase shifts.

\section{Conclusion}

In this paper, we demonstrate that third order polynomial regression can be used to align an acoustic image with an optical image with a fast calibration method that requires few measurements.
The results also show that SVD-PHAT can be efficiently used to generate the acoustic image in real-time with a reduction by a factor of $47$ in the number of computations compared to SRP-PHAT.
We plan to use this method in the future for multimodal segmentation and classification for robotic and other distant-talking audio-visual applications.

\clearpage

\bibliographystyle{IEEEbib}
\bibliography{refs}

\end{document}